\documentstyle[11pt,fleqn]{article}  
\input{psfig.tex}                
\title{Stationary Ballistic 'V' States   \\ for Preferred Motions
of Many Particles }
\author{A. Kwang-Hua Chu \thanks{The only author's  address after
2007-Aug. is : P.O. Box 30-15, Shanghai 200030, PR China.}} 
\date{P.O. Box 39, Tou-Di-Ban, Road XiHong, Urumqi 830000, PR China} 
\begin{document}
\maketitle
\begin{abstract}           
We use the discrete kinetic theory with the free-orientation
parameter being fixed ($\pi/4$) to derive the macroscopic velocity
field for many particles flowing through a microdomain. Our
results resemble qualitatively other hydrodynamical solutions. The
V-shaped velocity field changes as the dominant physical parameter
(Knudsen number) varies. We also briefly discuss the possible
mechanism due to the entropy production along the boundaries.

\vspace{3mm}

\noindent Keywords : \hspace*{1mm} Knudsen number, discrete
kinetic model, dilute gases.
\end{abstract}
\doublerulesep=6.8mm    
\baselineskip=6.8mm
\bibliographystyle{plain}
\section{Introduction}
Discrete kinetic theory [1-2] with the thermostat assumption or
diffuse scattering boundary condition [3] have been adopted to
{\it ad hoc} model the many-particle scattering situation along
the gas-surface interacting interface in a plane channel [1].
Specific orientations relevant to different rarefaction parameters
were identified therein [1]. Motivated by the recent interests in
the quantum Boltzmann approach [4] and the relevant studies [5-8],
we continue our previous studies [1] by examining the related
velocity and vorticity fields corresponding to those specific
$\theta$ and Kn we obtained and checking the special case
$\theta=\pi/4$. \newline Many interesting problems have been
successfully solved [9-17] by using the discrete kinetic theory.
Carleman (1957) developed 2-velocity models which are
defined by abstract properties in order to produce only
mathematical purposes. That model was not constructed on the basis
of detailed analysis of the collision mechanics [9]. Broadwell
(1964) devised a 6-velocity model to handle the simple shock wave
problem [10]. At first mathematicians have been interested in the
discrete Boltzmann equation  with  the hope to provide, thanks to
the relatively simpler structure of the equation as compared with
that of the full Boltzmann equation, stronger results than those
available for the Boltzmann equation or mathematical results
suitable to be technically generalized to the full Boltzmann
equation in the analysis of initial and initial-boundary value
problems. However, the analysis over recent years has shown that
this is not the case [5,7-8]. These have been reviewed considering
mainly the mathematical aspects of the initial and
initial-boundary value problems in order to provide a "more
detailed" analysis in a "more specialized" field. In fact the
consistency of the mathematical results depends on the structure
of the mathematical problems : in some cases it is possible to
obtain for the discrete Boltzmann equation "stronger" results than
the corresponding ones for the Boltzmann equation, and in other
case "weaker" results. Kawashima  has proved the global existence
of solutions to the initial-boundary value problems (I.-B.V.P.)
for the discrete Boltzmann Equation in the 1D-region $0<x<\infty$
or $0<x<d$ (cf [11]). \newline
Cornille obtained some transient or stationary family of solutions
for certain (fixed-orientation) discrete velocity models without
considering the boundary conditions [12]. 
\newline In this short paper, we plan to reconstruct the
macroscopic velocity field of many dilute particles by the
verified 4-velocity model [18] (the free orientation is fixed to
be $\theta =\pi/4$) considering a simple test problem : many
molecules or particles flowing along the bounded-plane channel and
finally reaching a steady state.
The verification of our approaches has been done in [1,6] (cf.
Chu), the argues about the differences between different discrete
velocity models included. For $\theta=\pi/4$ case, using a
completely different solving procedure, we obtained velocity
fields which have a V-shaped or chevron-structure.
\newline 
This short note is organized as follows. We introduce the general
orientation-free [5,6] 4-velocity model in Section 2, and simplify
it to a system of four equations for associated unknown functions.
The general boundary conditions will be briefly introduced, too.
Then, we define some macroscopic variables (like $u,v$) to suit
our interest which is to find a class of steady (and parallel)
non-boundary-driven solutions or flows for particles flowing along
a microslab with bounded (flat-plane) walls. The orientation will
be fixed as $\pi/4$ here when we solve the time-independent system
of equations with relevant boundary conditions for the test case.
As reported in [18], there will be no dispersion or absorption
when we implement the model with this orientation so that we can
resolve sharp velocity profiles. These kinds of solutions, $u$,
which collect the main results of the present paper, are given in
explicit form, and are functions of 1D coordinate : $y$ or $Y$ and
are also dependent on certain integration constants or parameters
due to the purely diffuse reflection boundary conditions. Finally,
we analyze the solutions (V-shaped fields) and make some physical
comments or discussions in comparison with the other flow-pattern
selection due to the relative orientation effect upon binary
encounter of many particles or unusual entropy production along
the confined boundaries.
\section{Formulations}
Considering a simple fluid of particles with mass $m$ and
cross-sectional area $\sigma$, the first step of the modelling
procedures consists in discretizing the velocity directions in a
finite number of unit vectors ${\bf i}_k$, $k=1,\cdots,p$. One or
more moduli are then associated to each direction. The ratio of
the moduli has, however, to be properly chosen, so that collisions
between particles with different velocity moduli are possible. For
one velocity moduli case, ${\bf u}_i$= $c {\bf i}_k$,
$k=1,\cdots,p$; $c\equiv c({\bf x}, t)$ in general. Normally $c$
is determined by the equilibrium distribution. \newline The
particles (hard-sphere) move in the whole space and collide by
simple elastic collisions locally in space. The mathematical model
is an evolutional equation for the number densities $N_i({\bf x},
t)$ linked to the finite set of velocities ${\bf u}_i$. We write a
balance equation for the number density of particles "i" in the
form
\begin{displaymath}
 [\frac{\partial }{\partial t} +{\bf u}_i \cdot \nabla ] N_i = G_i -L_i
\end{displaymath}
where $L_i$ and $G_i$ are the loss and the gain of the particles
"i" due to collisions. In case of binary collisions an exact
balance may be obtained, and is expressed with the transitional
probabilities and the number densities. This model has the
structure of a system of semi-linear partial differential
equations of hyperbolic type. Above equation could also be written
as
\begin{displaymath}
 \frac{\partial}{\partial t} N_i +{\bf u}_i \cdot \nabla N_i =
 \sum^R_{r=2} \sum_{I_r \in E_r} \sum_{J_r \in E_r} \delta (i,J_r,I_r)
  A_{I_r}^{J_r} N_{I_r},
\end{displaymath}
where $i=1,\cdots,p$; here, by definition, an $r$-collision ($r
\ge 2$) involves $r$ particles. $I_r$=($i_1,\cdots,i_r$), and
$J_r$ =($j_1,\cdots,j_r$) are two elements of $E_r$, which is the
set of $r$-not arranged numbers (considering the combinations
instead of the order they appear) taken in the set
$\{1,\cdots,p\}$. \newline A "transitional" probability denoted by
$A_{I_r}^{J_r}$ is associated to each $r$-collision $I_r
\rightarrow J_r$. In the case of binary collisions, this term
(also is called as the {\it transition rates}) is referred to the
collisions $({\bf u}_i,{\bf u}_j)$ $\leftrightarrow$ $({\bf
u}_k,{\bf u}_l)$, $i,j,k,l=1,\cdots,p$; and the number of
paired-outputs corresponding to a given paired-input is denoted by
$q$. $N_{I_r}$ denotes the product $N_{i_1} N_{i_2} \cdots
N_{i_r}$. \newline $\delta(i,J_r,I_r)$= $\delta(i,J_r)-$
$\delta(i,I_r)$ is the algebraic number of particles "i" created
through the collision $I_r \rightarrow J_r$. $\delta(i,I_r)$ is
(positive or zero) the number of indices $i$ present in the
$r$-set. 
If only nonlinear binary collisions are considered and considering
the evolution of $N_i$, we have
\begin{displaymath}
 \frac{\partial N_i}{\partial t}+ {\bf u}_i \cdot \nabla N_i
 =\sum^p_{j=1} \sum_{(k,l)} (A^{ij}_{kl} N_k N_l - A^{kl}_{ij}
 N_i N_j),  \hspace*{3mm} i=1,\cdots, p,
\end{displaymath}
where $(k,l)$ are admissible sets of collisions. We may then
define the right-hand-side of above equation as
\begin{displaymath}
 Q_i (N) =\frac{1}{2}\sum_{j,k,l} (A^{ij}_{kl} N_k N_l - A_{ij}^{kl}
 N_i N_j),
\end{displaymath}
with $i \in$ $\Lambda$ =$\{1,\cdots,p\}$, and the summation is
taken over all $j,k,l \in \Lambda$, where $A_{kl}^{ij}$ are
nonnegative constants satisfying \hspace*{3mm}
  $A_{kl}^{ji}=A_{kl}^{ij}=A_{lk}^{ij}$ :
  $\mbox{indistinguishability of the particles in collision}$,
 $A_{kl}^{ij} ({\bf u}_i +{\bf u}_j -{\bf u}_k -{\bf u}_l )=0$ :
 $\mbox{conservation of
 momentum in collision}$,
 $A_{kl}^{ij}=A_{ij}^{kl}$ : {microreversibility condition}.
The conditions defined for the discrete velocity above requires
that elastic, binary collisions, such that momentum and energy are
preserved
 ${\bf u}_i +{\bf u}_j = {\bf u}_k +{\bf u}_l$,
 $|{\bf u}_i|^2 +|{\bf u}_j|^2 = |{\bf u}_k|^2 +|{\bf u}_l|^2$,
are possible for $1\le i,j,k,l\le p$. \newline The collision
operator is now simply obtained by joining $A_{ij}^{kl}$ to the
corresponding transition probability densities $a_{ij}^{kl}$
through $ A_{ij}^{kl}$ =$ S|{\bf u}_i-{\bf u}_{j}|$ $a_{ij}^{kl}$,
where,
\begin{displaymath}
 a_{ij}^{kl} \ge 0 , \hspace*{12mm} \sum^p_{k,l=1}  a_{ij}^{kl}=1 ,
 \hspace*{3mm} \forall i,j=1,\cdots,p ;
\end{displaymath}
with $S$ being the effective collisional cross-section. If all $q$
($p=2q$) outputs are assumed to be equally probable, then
$a_{ij}^{kl}$=$1/q$ for all $k$ and $l$, otherwise $a_{ij}^{kl}$=
0. The term $ S|{\bf u}_i-{\bf u}_{j}| dt$ is the volume spanned
by the particle with ${\bf u}_i$ in the relative motion w.r.t. the
particle with ${\bf u}_j$ in the time interval $dt$. Therefore, $
S|{\bf u}_i$ $-{\bf u}_{j}| N_j$ is the number of $j$-particles
involved by the collision in unit time. Collisions which satisfy
the conservation and reversibility conditions which have been
stated above are defined as admissible collisions.
\newline
The discrete kinetic equations then [1,5,7,18] assume the
following form
\begin{displaymath}
 \frac{\partial N_i}{\partial t}+c [\cos(\theta+(i-1)*\pi/q)\frac{\partial
 N_i}{\partial x}+\sin(\theta+(i-1)*\pi/q)\frac{\partial N_i}{\partial y}]=
\frac{2 c S}{q}{\sum^q_{j=1}}_{j\ne i}(N_j N_{j+q}-
\end{displaymath}
\begin{equation}
 N_i N_{i+q}) \hspace*{2mm} \mbox{or} \hspace*{2mm} =\frac{2 c S}{q}
 \sum^{q-1}_{l=1} (N_{i+l} N_{i+l+q}-N_i N_{i+q} ); \hspace*{3mm}
 i=1,\cdots, 2q,
\end{equation}
where $\theta$ is the free orientation starting from the positive
$x-$axis to the $u_1$ direction [1,18], $N_i=N_{i+2q}$ are unknown
functions,
and 
$c$ is a reference velocity modulus.\newline 
According to [13], for the $2q$-velocity model that is $q \ge 3$,
there are more {\it collision invariants}
than the physical ones or conservation laws which are
corresponding to the number of macroscopic variables (in 2D. there
are only 4, i.e., one mass, two momenta, one energy). That's to
say, there are unphysical or {\it spurious} invariants or
macroscopic variables for $q \ge 3$ models (which could be,
however, well handled by adding multiple collisions [13]). Thus,
we plan to use only the orientation-free 4-velocity model for our
test-case problem below.
\newline

\setlength{\unitlength}{0.8mm}
\begin{picture}(60,50)(-35,-5)
\thicklines \put(20,20){\vector(1,2){10}}
\put(23,42){\makebox(0,0)[bl]{\large {\bf u$_1$}}}
\put(20,20){\vector(-2,1){20}} \put(12,0){\makebox(0,0)[bl]{\large
{\bf u$_3$}}} \put(20,20){\vector(-1,-2){10}}
\put(40,6){\makebox(0,0)[bl]{\large {\bf u$_4$}}}
\put(20,20){\vector(2,-1){20}} \put(4,32){\makebox(0,0)[bl]{\large
{\bf u$_2$}}} \thinlines \put(0,10){\vector(2,1){20}} \thinlines
\put(0,10){\vector(1,1){30}}          
\thinlines
\put(0,10){\vector(0,2){20}}          
\thinlines \put(0,10){\vector(1,-1){10}} \thinlines
\put(20,20){\line(2,0){20}}           
\put(20,20){\oval(5,5)[tr]}
\put(20.7,22.5){\makebox(0,0){$\theta$}}
\put(21,16){\makebox(0,0){{\bf G}}} \put(-2,11){\makebox(0,0){{\bf
O}}} \thinlines \put(0,10){\vector(4,0){40}}
\put(-2,-10){\makebox(0,0)[bl]{\small  Fig. 1 \hspace*{2mm}
Reference frame for the 4-velocity model with $\theta=\pi/4$
here.}}
\end{picture}
\subsection{Boundary Conditions}
We use purely diffuse reflection boundary condition [1,3,15-16]
here, which means properties of the reflected particles are
independent of their properties before the impact. In other words,
the re-emitted stream has completely lost its memory of the
incoming stream, except for the conservation of the number of
particles. Moreover, we impose the following conditions : the
particles are in Maxwellian equilibrium with the wall ("the wall
locally behaves as a thermostat", i.e., the particles reflect
after they have been in thermodynamic equilibrium with the
wall-temperature) satisfies $N_i ({\bf r},t)$=$\gamma_i ({\bf
r},t) N_{wi} ({\bf r},t)$, where $\gamma_i$ expresses the
accomodation of the particles to the wall quantities, and $N_{wi}$
is the discrete Maxwellian densities for the 'i'-direction set of
particles. That is, we have
\begin{equation}
 |{\bf u}_j \cdot {\bf n}| N_{wj}=\sum_{i \in I} B_{ij} |{\bf u}_i \cdot
 {\bf n}| N_{wi}, \hspace*{3mm}  j\in R,  \hspace*{3mm} B_{ij} \ge 0,
 \hspace*{3mm} \sum_{j\in R} B_{ij} =1 ;
\end{equation}
with $I=\{i,({\bf u}_j-{\bf u}_w)$ $\cdot {\bf n} <0 \}$ related
to the impinging set of particles, $R=\{j,({\bf u}_j-{\bf u}_w)$
$\cdot {\bf n}$ $>0 \}$ related to the emerging set of particles,
${\bf n}$ is the outer normal, ${\bf u}_w$ is the wall velocity.
\subsection{Flows in a Plane Channel}
We firstly define the related macroscopic variables $n$ = $N_1
+N_2 +N_3 +N_4$, $n U$ =$ c (\alpha N_1- \beta N_2-\alpha
N_3+\beta N_4)$, $n V$ = $c (\beta N_1+\alpha N_2$ -$\beta N_3-
\alpha N_4)$, (the latter two are the momentum flux along $x$- and
$y$-directions) with $\rho=n\,m$, $m$ is the mass of the molecule,
$\rho$ is the macroscopic density of the gas. Then, set $n_i =N_i
/ n $, $i=1,2,3,4$; and then use non-dimensional $u=U/ c$,
$v=V/c$, $Y =y/d$, where $c$ may be related to the external
forcing [1,19]. $d$ is the full channel width. $y=0$ is along the
center-line.
\newline The geometry of a 2D problem we shall consider is a kind
of microchannels with bounded flat-plane walls which are separated
apart by a width $d$. Particles (driven by an external constant
forcing initially) flowing along this channel will finally reach a
fully developed state (steady state and $\partial u/\partial x=0$,
$v=0$). \newline We derive the solutions with $\alpha (\equiv\cos
\theta)=\beta(\equiv\sin \theta) =\sqrt{2}/2$ case here. The
algorithm is different from those previously reported, we must
solve the independent number density respectively then combine
them into macroscopic ones since the original macroscopic equation
is singular (cf. equations in [6] by Chu). Meanwhile, from the
preliminary results reported in Ref. [1,18], it seems, for the
case of $\theta=\pi/4$, 4-velocity model will give completely
different dispersion relations for the thermodynamic checking of
the perturbed Maxwellian equilibrium state. There will be no
dispersion or absorption for this particular case.
\newline
The governing equations (1), for the assumptions prescribed above,
now become
\begin{equation}
 \frac{d n_1}{d Y}=-\frac{d n_2}{d Y}=-\frac{d n_3}{d Y}=
 \frac{d n_4}{d Y}=\frac{\sqrt{2}}{4 \mbox{Kn}} (1-2 a)=\frac{\sqrt{2}}{\mbox{Kn}}
 (n_2 n_4 -n_1 n_3 ) ,
\end{equation}
here, $n_3 =a-n_1$, $n_2=1/2 -n_1$, $n_4=1/2-a+n_1$; Kn$=1/(d S
n)$ is the Knudsen number. The diffuse reflection boundary
conditions become :
\begin{equation}
  N_{w2} N_1 =N_{w1} N_2 , \hspace*{6mm}
  \beta N_1 +\alpha N_2 -\beta N_3 -\alpha N_4 =0 ,
\end{equation}
it means (i) the Maxwellian equilibrium at the walls dominates,
(ii) no penetration occurs across the wall. The discrete
Maxwellian densities $N_{wi}$ at the wall, as derived before
(please see the detailed references in Refs. [14-16]), are
\begin{equation}
 N_{wi}=(n/4) \{ 1+(2/c^2){\bf u_w}\cdot{\bf u}_i+(-1)^i [({\bf u}_w \cdot
 {\bf u}_2)^2-({\bf u}_w \cdot{\bf u}_1)^2](1/c^4) \}.
\end{equation}
Here, boundary conditions are, as ${\bf u}_w=0$ (the walls are
static and fixed) and by assuming the symmetry,
\begin{equation}
 n_1 =B_{31} n_3 +B_{41} n_4 , \hspace*{3mm} \mbox{at } Y=-1/2,
 \hspace*{6mm} n_3 =B_{13} n_1 +B_{23} n_2 , \hspace*{3mm} \mbox{at }
 Y=1/2,
\end{equation}
with the discrete Maxwellians $N_{wi} |_{\pm}=1/4$. Integration of
Eq. (3) gives
\begin{displaymath}
 n_1 =\frac{\sqrt{2}(1-2 a)}{4 \mbox{Kn}}Y +b.
\end{displaymath}
Now, set $A=1/(4\sqrt{2}$ Kn), so we get from above equations to
solve for $a,b$ :
\begin{equation}
 [2 A (1+B_{31}-B_{41} )-B_{31}+B_{41} ] a +(1+B_{31}-B_{41} ) b=
 \frac{B_{41}}{2}+(1+B_{31}-B_{41} ) A  ,
\end{equation}
\begin{equation}
 [1+ 2 A (1+B_{13}-B_{23})] a +(B_{23}-B_{13}-1) b= \frac{B_{23}}{2}+ A
 (1+B_{13}-B_{23} )  , \hspace*{6mm} \mbox{or}
\end{equation}
\begin{equation}
 C \, a +D \, b = G ,  \hspace*{12mm} E \, a +F \, b = H  .
\end{equation}
After manipulations, we have
\begin{equation}
 a=\frac{F\,G-D\,H}{C\,F-D\,E} ,\hspace*{10mm}
 b=\frac{C\,H-E\,G}{C\,F-D\,E} ;
\end{equation}
where $C\, F-D\, E =4 A (B_{23}-B_{13}-1)+(B_{23}-B_{13})
(B_{41}-B_{31})-1$, and
 $F G-D H= (B_{23}-B_{13}-1)[A(1+B_{31}-B_{41})+{B_{41}}/{2}]+(1+B_{31}-B_{41})
 [A(B_{23}-B_{13}-1)-$${B_{23}}/{2}]$,
 $C H-E G=A [(B_{23}-1)(1-B_{41})-B_{31}(2+B_{13}-2B_{31})]+
   $$[{(B_{23}-1)B_{41}-B_{23}B_{31}}]/{2}$.
\newline Since $n U/c=\sqrt{2}/2 (N_1 -N_2 -N_3 +N_4)$, so we have a
family of (particular) flow field in terms of the macroscopic
velocity
\begin{equation}
 u =\sqrt{2} (2 n_1 - a)=\frac{1-2 a}{\mbox{Kn}} Y+\sqrt{2} (2 b-a) .
\end{equation}
\section{Results and Discussions}
This class of solution $u$ obtained by fixing the orientation to
be $\pi/4$ is in general different from those reported in Ref. [6]
by Chu. Note that, for one extreme case of boundary conditions as
mentioned in Eq. (2) : $B_{31}=B_{41}$, and $B_{13}=B_{23}$; we
have
\begin{equation}
 a=\frac{2A+(B_{41}+B_{23})/2}{1+4A}, \hspace*{12mm}
 b=\frac{A(1+B_{41}-B_{23})+B_{41} /2}{1+4A}.
\end{equation}
We can easily observe that, from equation (11), that $u=0$
everywhere for all Knudsen numbers (Kn). There is no macroscopic
flow [20] for many particles once the boundary conditions are
selected above.
\newline Otherwise, the velocity field (from equation (11)) as shown in
figure 2 is qualitatively similar to the V-shaped or {\it
chevron}-like structure or pattern [21] reported before in other
physical systems. The velocity field is tuned mainly by the Kn and
weakly by $a$ and $b$ with the latter due to the boundary
conditions.
We note that $a$ might depend on the physical properties of fluids
and the geometry of the solid-wall as it comes from the gas-solid
interaction or reflection. The flow-pattern selection mechanism is
yet open to the best knowledge of the authors but might be
partially linked to that reported in Ref. [1,6] (cf. Chu) since
there will be an essential singularity when integrating equation
(1) for $\theta=\pi/4$ case. In short, as Kn increases, the
chevron front becomes more flat.
\newline The macroscopic vorticity $\omega$ (or the mean shear) could be obtained by
noting
\begin{equation}
 \omega=\frac{d u}{dY}= \frac{1-2a}{\mbox{Kn}} +\frac{d [\sqrt{2} (2
 b-a)]}{d Y}, \hspace*{12mm}
\end{equation}
with
\begin{displaymath}
 a=\frac{2A (1+B_{31}-B_{41})(B_{23}-B_{13}-1)+[2
 B_{41}B_{23}-B_{41}(B_{13}+1)-B_{23}(1+B_{31})]/2}{4A(B_{23}-B_{13}-1)+
 (B_{23}-B_{13})(B_{41}-B{31})-1},
\end{displaymath}
where the last term of equation (13) is generally zero. Once the
Knudsen number (Kn; a kind of rarefaction measure for
many-particles interactions or collisions) is fixed, the vorticity
is a constant with the related $B_{ij}$ subjected to the
constraint in equation (2). In fact, $B_{ij}$ should depend on the
detailed interactions of the gas-solid interface, like a kind of
(known) molecules colliding with specific walls made of (already)
specified material. It is bounded above but difficult to be fixed
even for specific model and boundary value problem [11].\newline
Our results for the vorticity field, at least, qualitatively
matches with the hydrodynamic two-dimensional solution
\cite{Fluid:Landau} when the weakly compressible (incompressible)
particles flow along a static flat-plane channel and finally reach
a fully developed state even though the particles are initially
driven by a constant pressure-gradient or unit forcing.
Interestingly, similar sharp flow fields of  solitary wave profiles (the highest one, cf.
Figs. 9 and 10 by Wu {\it et al.} in [22]) and constant-V vortex
was reported recently [22] in other physical system dealing with
confined flow transports.
\newline
To further interpret the mechanism, we propose that the
complicated rate of entropy production along the boundaries (cf.
[23-24]) might favor the smearing of viscous diffusion (toward the
away-from-wall regions) so that the sharp and strange pattern
could form and then there is no significant smoothing of the
profiles along the cross-section. \newline Note that, the approach
here : firstly tracing or obtaining (solving the corresponding
equation in (3)) each individual discrete number density ($n_i$)
then by summing up the corresponding projection to obtain $u$, is
different from that in [6] (by Chu) : directly construct the
macroscopic solutions from the relevant governing equation for
macroscopic variables ($u$). The boundary treatment which is
relevant to the entropy production there is thus entirely
different. The corresponding non-equilibrium states (due to
different rates of entropy production and their decay) approaching
to the final equilibrium states which are used as our boundary
conditions might then be different. In fact, as we noticed,  the
argument raised in [24] could be applicable to present approach as
evidenced in the boundary operator as expressed in equation (2)
(could be represented as similar divergence form). Otherwise, if
our interpretations don't work, there might exist other unknown
mechanism which need our further works.
\newline {\bf Acknowledgements.}
The author is partially supported by the China Post-Dr. Science
Foundation under Grant No. 1999-17.

\newpage

\pagestyle{myheadings} \topmargin=-20mm \textwidth=16cm
\textheight=28.5cm \oddsidemargin=-6mm

\psfig{file=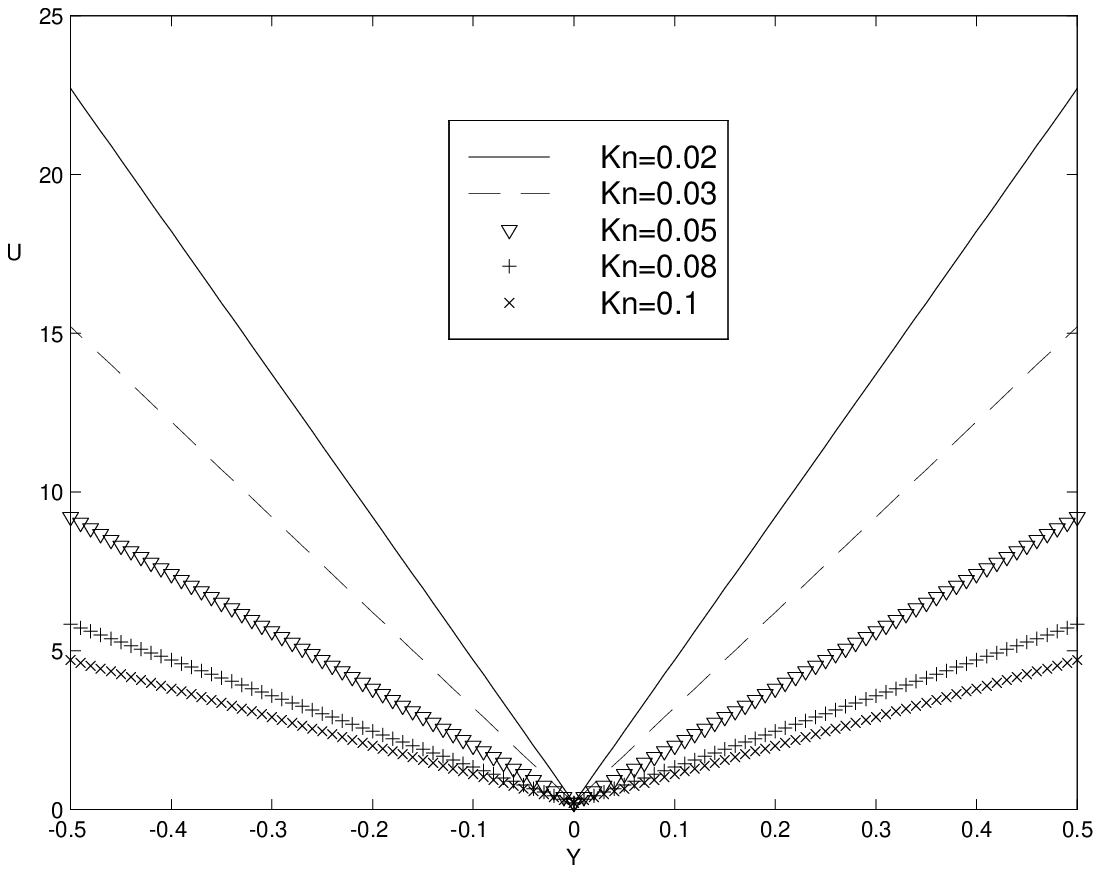,bbllx=0.6cm,bblly=6.8cm,bburx=18.5cm,bbury=27cm,rheight=10cm,rwidth=10cm,clip=}

\vspace{10mm}
\psfig{file=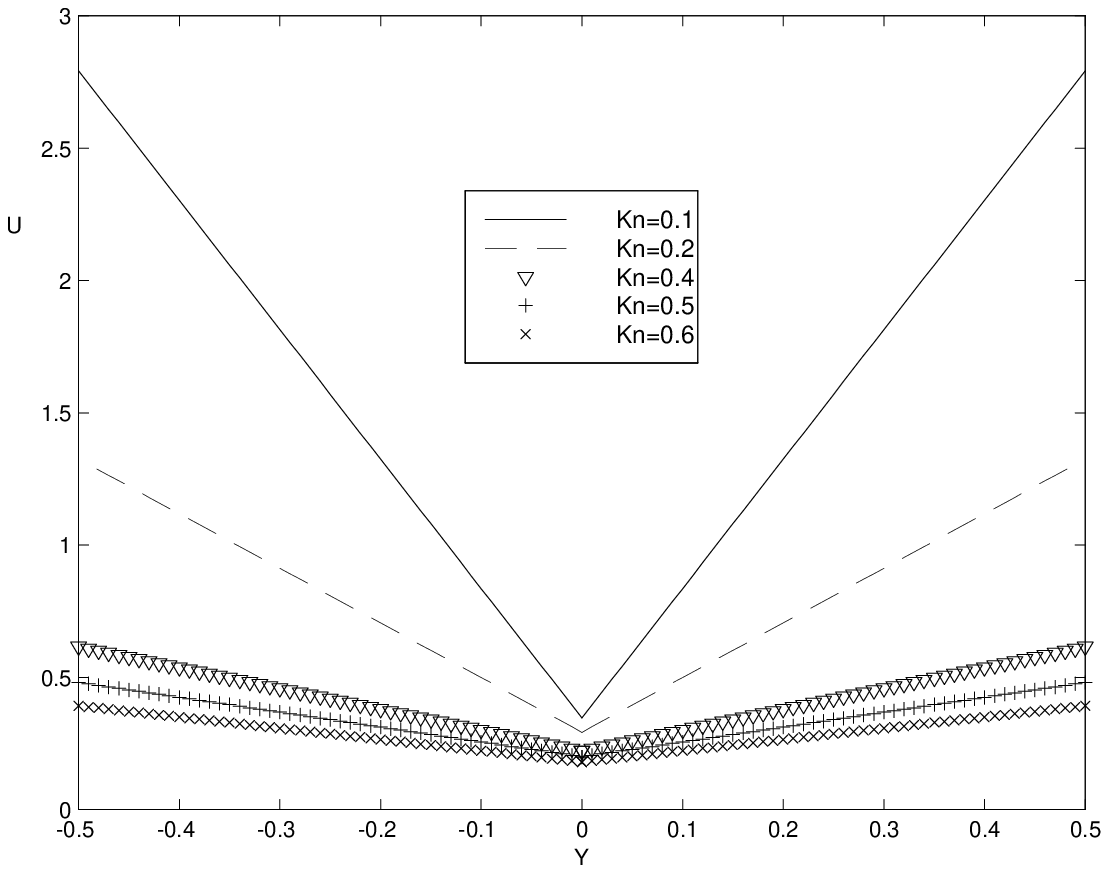,bbllx=0.6cm,bblly=6.8cm,bburx=18.5cm,bbury=27cm,rheight=10cm,rwidth=10cm,clip=}

\vspace{16mm}
\begin{figure}[h]
 \hspace*{10mm} Fig. 2 (a),(b) \hspace*{3mm}
Rarefaction effects (Kn) on the velocity field $u$ or the
\newline \hspace*{11mm} V-shaped or chevron-like structure. Kn$=1/(d\,
S\, n)$ is the Knudsen  number. \newline \hspace*{11mm} $S$ is the
effective collision cross-section. $n$ is the number density of
particles.

\end{figure}

\end{document}